\documentclass[twocolumn,secnumarabic,amssymb, showpacs, nobibnotes, aps, prd]{revtex4-1}

\usepackage{graphicx}
\usepackage{subfigure}
\usepackage{ae} %pdf-Schriftarten
\usepackage{float}  % allows to use the [H] parameter for includegraphics- placed the picture exactly at this place.
\usepackage{amsmath}
\usepackage{color}

\begin{document}

\title{Anatomy of inertial magnons in ferromagnetic nanostructures}

\author{Alexey M.~ Lomonosov$^{1}$}
\email[]{lom@kapella.gpi.ru}

\author{Vasily V.~ Temnov$^{2,3}$}
\email[]{vasily.temnov@univ-lemans.fr}

\author{Jean-Eric Wegrowe$^3$}
\email[]{jean-eric.wegrowe@polytechnique.edu}

\affiliation{$^1$Prokhorov General Physics Institute of the Russian Academy of Sciences, 119991, Moscow, Russia}

\affiliation{$^2$Institut des Mol\'{e}cules et Mat\'eriaux du Mans, UMR CNRS 6283, Le Mans Universit\'e, 72085 Le Mans, France}
\affiliation{$^3$LSI, Ecole Polytechnique, CEA/DRF/IRAMIS, CNRS, Institut Polytechnique de Paris, F-91128, Palaiseau, France}

\date{\today}

\begin{abstract}
 We analyze dispersion relations of magnons in ferromagnetic nanostructures with uniaxial anisotropy taking into account inertial terms, i.e. magnetic nutation. Inertial effects are parametrized by damping-independent parameter $\beta$, which allows for an unambiguous discrimination of inertial effects from Gilbert damping parameter $\alpha$. The analysis of magnon dispersion relation shows its two branches are modified by the inertial effect, albeit in different ways. The upper nutation branch starts at $\omega=1/ \beta$, the lower  branch coincides with FMR in the long-wavelength limit and deviates from the zero-inertia parabolic dependence  $\simeq\omega_{FMR}+Dk^2$ of the exchange magnon. Taking a realistic experimental geometry of magnetic thin films, nanowires and nanodiscs, magnon eigenfrequencies, eigenvectors and $Q$-factors are found to depend on the shape anisotropy. The possibility of phase-matched magneto-elastic excitation of nutation magnons is discussed and the condition was found to depend on $\beta$, exchange stiffness $D$ and the acoustic velocity.

\end{abstract}

\pacs{Valid PACS appear here}


\maketitle
%\tableofcontents

\section{Introduction}
After the first description of the dynamics of the magnetization by  Landau and Lifshitz \cite{Landau1935}, Gilbert proposed an equation that contains a correction due to the precessional damping \cite{gilbert1956formulation,gilbert2004phenomenological}. Since then, the so-called Landau-Lifshitz-Gilbert (LLG) equation is known to give an excellent  description of the dynamics of the magnetization, including ferromagnetic resonance (FMR) and magnetostatic waves \cite{damon1961magnetostatic,farle1998ferromagnetic}, as well as the magnetization reversal \cite{thevenard2013irreversible,Vlasov2020}. Ferromagnetic resonance and time-resolved magnetization measurements allow its spatially homogeneous precession ($k=0$) but also non-uniform modes of the magnetization precession ($k\neq 0$, where $k$ is the wave vector of spin waves) to be measured \cite{van2002all,kruglyak2010magnonics,razdolski2017nanoscale}. During the last decades, these techniques have been advanced in the context of ultrafast demagnetization dynamics \cite{beaurepaire1996ultrafast,kirilyuk2010ultrafast} that paved the way for the description of new physics at the sub-picosecond regime. High-frequency resonant modes of exchange magnons have been measured with ultrafast time-resolved optical techniques \cite{van2002all,salikhov2019gilbert,razdolski2017nanoscale}. Therefore, the validity of the LLG equations has been confirmed down to the picosecond time scale and below. 

However, limitations of LLG equations has been established in the stochastic derivation performed by W. F. Brown in a famous paper published in 1963 \cite{brown1963thermal}. This limit is due to the hypothesis that the typical time scales of magnetization dynamics are much longer than those of other degrees of freedom forming the dissipative environment. In analogy to the common description of the diffusion process of a Brownian particle, the inertial (momentum) degrees of freedom are supposed to relax much faster than its spatial coordinate. This means that the degrees of freedom related to the linear momentum (in the case of the usual diffusion equation), or to the angular momentum (in the case of the magnetization) are included into the heat bath. As a consequence, the inertial terms do not explicitly appear in the equations, but are considered to be part of the damping term \cite{wegrowe2012magnetization}.

The possibility of measuring the contribution to inertial degrees of freedom led to a generalization of the LLG equation with an additional term, incorporating the second time-derivative of magnetization:

\begin{equation}
\dot{\mathbf{m}}=-\gamma \mathbf{m}\times \mathbf{H_{eff}}+\alpha \mathbf{m}\times \dot{\mathbf{m}} +\beta \mathbf{m}\times \ddot{\mathbf{m}}\,,
\label{eq:ILLG}
\end{equation}

where $\mathbf m = \mathbf M/M_s$ is the unit magnetization vector that gives the direction of the magnetization at each point (and $M_s$ is the modulus of the magnetization, which is constant), $\gamma=\gamma_0 \mu_0$ is the gyromagnetic ratio, $\alpha$ stands for the Gilbert damping. Inertial effects are characterized by the parameter $\beta$, which is introduced in a phenomenological way, i.e. independent on $\alpha$ and $\gamma$ \footnote{In a majority of the published reports on the magnetic inertia, the coefficient used is the relaxation time $\tau$ of the inertial degrees of freedom $\tau = \beta / \alpha$, which has a clear physical meaning. However, only the parameter $\beta$ is intrinsic, i.e. proper to the material. Indeed, in the framework of the mechanical analogy, $\beta$ is defined by the first and second inertial moment $I_1 = I_2$ see Ref.~\cite{wegrowe2012magnetization}.}. 
This generalized LLG equation has been derived in the framework of different and independent theoretical contexts \cite{wegrowe2000thermokinetic,fahnle2011generalized,fahnle2013erratum,ciornei2011magnetization,wegrowe2012magnetization, bhattacharjee2012atomistic,kikuchi2015spin,wegrowe2016magnetic,sayad2016inertia,thonig2017magnetic,mondal2017relativistic,bajpai2019time,fahnle2019comparison,mondal2020dynamics,giordano2020derivation,mondal2021nutation,titov2021inertial}; its solutions have been studied in a series of publications \cite{olive2012beyond,olive2015deviation,bottcher2012significance,cherkasskii2020nutation}. The main consequence of inertia for the uniform magnetization (magnon with the wavevector $k=0$) is the existence of nutation oscillations that are superimposed to the precession. This leads to an appearance of the second resonance peak at a higher frequency in FMR spectra. The direct measurement of nutation has been reported recently \cite{neeraj2021inertial,li2015inertial}. 

The goal of the present report is to study the consequences of these inertial effects on the  exchange magnons (i.e. $k\ne 0$ modes), in the perspective of experimental studies. Magnons are defined as linear magnetic excitations propagating in ferromagnets at the micromagnetic limit. This work completes the first description published in 2015, Section IV of the remarkable work of Toru Kikuchi and Gen Tatara \cite{kikuchi2015spin}, and independently reconsidered by Makhfudz et al. in 2020 \cite{makhfudz2020nutation}. 

The paper is organized as follows. Section II presents the derivation of the linear magnetic excitations deduced from \eqref{eq:ILLG}. Section III describes the dispersion relation in a simple case of zero dipolar field (spherical symmetry). The first consequence of the inertia is that the dispersion relation splits in two branches: the lower one $s_1{\rm exp}(ikz-i\omega_1t)$ (\textit{FMR magnons}) and the upper one for $s_2{\rm exp}(ikz-i\omega_2t)$ (\textit{nutation magnons}). The second consequence is that the quality factor $Q$ increases with the $k$-vector. Section IV generalizes the description to the case of a uniaxial anisotropy quantified by the dimensionless (shape) anisotropy parameter $\xi$. In the anisotropic case the trajectories of both FMR magnons and nutation magnons become elliptical and rotating in opposite directions at each point in space. For a given $k$-vector the magnetization vector corresponding to a superposition of both magnons draws a typical trochoidal trajectory  (see Fig.~ \ref{fig:Arrangement}). Section V discusses the conditions for phase-matched excitation the nutation magnons by co-propagating longitudinal acoustic phonons, illustrated by the material parameters for Gd-doped Permalloy thin films \cite{salikhov2019gilbert}.

\section{Exchange magnons in ferromagnetic thin films with magnetic inertia}

\begin{figure}
	{\footnotesize{} \centering
 \includegraphics[width=1.0\columnwidth]{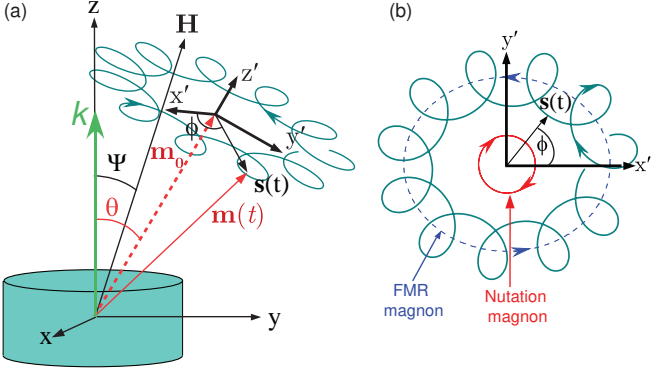}
 \caption{\label{fig:Arrangement} (a) Inertial magnons propagating in ferromagnetic nanostructures with wavevector $k$ along the $z$ direction under an external magnetic field $\mathbf{H}$ result in complex magnetization dynamics. (b) They can be decomposed in FMR magnon and nutation magnon precessing in opposite directions on elliptical trajectories at different frequencies, giving rise to a characteristic {\it flower-shaped} trajectory.}}
\end{figure}

We start with the LLG equation for unit magnetization vector $\mathbf m$ with an effective field $\mathbf{H}_{eff}$, which includes exchange interactions with stiffness {\it D}, an external field $\mathbf{H}=(H_{x}, 0, H_{z})$ and a demagnetizing field induced by the shape anisotropy $\mathbf{H_{d}}=-M_{S}\widehat{N}\mathbf{m}$. The demagnetization tensor $\widehat{N}$ depends on the specific shape of the ferromagnetic sample. Hereafter we assume the diagonal form of $\widehat{N}$ with diagonal elements $N_x$, $N_y$ and $N_z$. Damping of the magnetization dynamics is described by the conventional Gilbert term with parameter $\alpha$. In addition to the conventional LLG equation we take into account the inertial effect characterized by the independent parameter $\beta$. Then the inertial LLG equation (ILLG) takes the form of Eq.(\ref{eq:ILLG}) with $\mathbf{H_{eff}}=\mathbf{H}+D\Delta \mathbf{m}+\mathbf{H_{d}}$. 

The coordinate system was chosen such that the external field lies in the $\mathit{y}=0$ plane, as is shown in Fig.~\ref{fig:Arrangement}. The material is assumed to be magnetically isotropic, so that the unperturbed magnetization vector also lies in the $\mathit{y}=0$ plane. We seek for time- and space-dependent solutions in the form $\mathbf{m}=\mathbf{m_{0}+\mathbf{s}\left (z,t \right )}$ with spin-wave solutions

\begin{equation}
    \mathbf{s}\left ( z,t \right )=(s_x,s_y,s_z)\exp \left ( ikz-i\omega t \right )
\label{eq:s}
\end{equation}

propagating as plane waves with a real wave vector $k$ along the $\mathit z$-axis, see Fig.~\ref{fig:Arrangement}.
Substitution $\mathbf{m(z,t)}$ into equation~(\ref{eq:ILLG}) and its linearization with respect to small perturbations ${\it s_x}, {\it s_y}, {\it s_z}$ results in a homogeneous system of three linear equations:

\begin{equation}
\widehat{A}\begin{pmatrix}
s_x\\ 
s_y\\ 
s_z
\end{pmatrix}=0 \label{eq:system}
\end{equation}

where the matrix $\mathit A$ is given by

\begin{equation}
\left (\begin{array}{ccc}
-i\omega &  A_{12}(\omega,k)  & 0 \\
A_{21}(\omega,k)  & -i\omega & A_{23}(\omega,k) 	 \\
 0 & A_{32}(\omega,k)   & 	-i\omega \\	
 \end{array} \right )
 \label{eq:determ}
 \end{equation}
 
with coefficients $A_{ij}(\omega,k)$ defined as:

\begin{eqnarray}
   \nonumber
   A_{12}&=&m_z(\gamma D k^2+\gamma M_S \xi_{yz}-i\alpha  \omega-\beta \omega^2)+\gamma H_z \\
   \nonumber
   A_{21}&=& - m_z(\gamma D k^2+\gamma M_S \xi_{xz}-i\alpha  \omega-\beta \omega^2)+\gamma H_z\\
    \label{eq:A_ij}
   A_{23}&=& m_x(\gamma D k^2+\gamma M_S \xi_{zx}-i\alpha  \omega-\beta \omega^2)+\gamma H_x\\
    \nonumber
   A_{32}&=& -m_x(\gamma D k^2+\gamma M_S \xi_{yx}-i\alpha  \omega-\beta \omega^2)-\gamma H_x
\end{eqnarray} 

 where coefficients $\xi_{ij}=N_i-N_j$ characterize the shape anisotropy. The condition for the nontrivial solution of the homogeneous system (\ref{eq:system}) to exist, i.e. $\det A=0$, gives rise to the secular equation 
 
\begin{equation}
\omega^2+A_{12}(\omega,k)A_{21}(\omega,k) +A_{23}(\omega,k) A_{32}(\omega,k)=0
\label{eq:secular}
\end{equation}

which is used to calculate the spin wave dispersion relation $\omega(\mathit k)$ for different shapes/symmetries, {\it i.e.} characterized by different types of the $\widehat{N}$ tensor.

\section{Inertial exchange magnons in samples with spherical symmetry}
Examples of such symmetry are infinite homogeneous isotropic ferromagnetic media, or any spherical body. In these cases the demagnetization tensor $\widehat{N}$ is diagonal with all nonzero elements equal $1/3$, so that its contribution to the magnetization dynamics \eqref{eq:ILLG} and correspondingly to the wave matrix components \eqref{eq:A_ij} vanishes. The secular equation \eqref{eq:secular} takes a concise form:

\begin{eqnarray}
\left(\gamma H + \gamma D{k^2} - \beta\omega^2 - i\alpha\omega  + \omega  \right)\times\label{eq:unbounded det}\\ \times\left( \gamma H + \gamma Dk^2 - \beta\omega^2 - i\alpha\omega - \omega  \right) = 0\nonumber\,.
\end{eqnarray}

Due to the symmetry of $\widehat{N}$, equation \eqref{eq:unbounded det}, and hence all its roots, remains independent on the direction of $\mathbf H$ and the equilibrium magnetization $\mathbf {m}_0$ with respect to the wave propagation direction along the $\mathit z$-axis. For each positive wavenumber $\mathit k$, the determinant {\eqref{eq:unbounded det}} is solved for $\omega$. Given that the presumed solution has a form $\sim\exp \left ( ikz-i\omega t \right )$, positive $\omega$ designates the waves travelling in the positive direction. The two positive roots corresponding to the first parenthesis in \eqref{eq:unbounded det} have the following forms:

\begin{equation}
\omega _1 = \frac{1}{{2\beta }}\left( { - 1 - i\alpha  + \sqrt {4\gamma \beta (D{k^2} + H) + {(1 + i\alpha )^2}} } \right)
\label{eq:exact omega1}
\end{equation}

\begin{equation}
\omega_2 = \frac{1}{{2\beta }}\left( {1-i\alpha  + \sqrt {4\gamma \beta (D{k^2} + H) + {(1-i\alpha )}^2} } \right)
\label{eq:exact omega2}
\end{equation}

The first root is the lower magnon branch or precession, slightly modified by the inertial term and the second one exhibits the inertial magnon branch or nutation. It is convenient to split these roots into real and imaginary parts: $\omega_{1,2}=\omega_{1,2}^{\prime}+i\omega_{1,2}^{\prime\prime}$.
Taylor series approximation of those roots assuming the smallness of $\gamma \beta Dk^{2}, \gamma \beta H, \alpha \ll 1$ results in the following expressions for their real parts:

\begin{eqnarray}
\omega _1^{\prime} &\approx& \gamma [D{k^2} + H - 2\beta\gamma HD{k^2} + ...]\label{eq:approx omega1}\\
\omega _2^{\prime} &\approx& \frac{1}{\beta}+\omega_1^{\prime}\label{eq:approx omega2}  
\end{eqnarray}

\begin{figure}[htb!]
	{\footnotesize{} \centering
 \includegraphics[width=1.0\columnwidth]{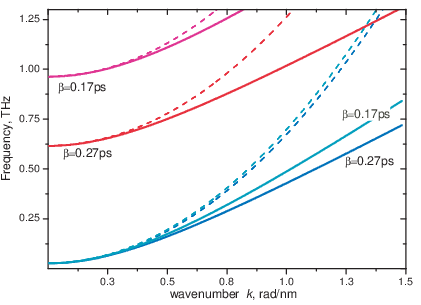} 
 \includegraphics[width=1.0\columnwidth]{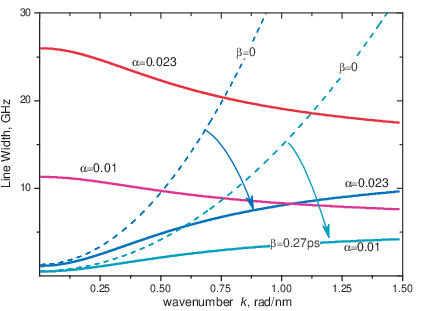}
 \caption{\label{fig:omega_real} (a) Dispersion of the inertial magnon (red, magenta) and conventional magnon (blue,green). Dashed curves shows the approximations by \eqref{eq:approx omega1} and \eqref{eq:approx omega2}. (b) corresponding line widths. In order to indicate the effect of inertia on the precession, dashed curves show the line width without inertia $\beta=0$.}
 }
\end{figure}

In this approximation the nutation magnon branch is simply shifted by $1/\beta$ with respect to FMR magnon branch. The validity of this approximation is illustrated in Fig.~\ref{fig:omega_real}, where the exact roots given by \eqref{eq:exact omega1} and \eqref{eq:exact omega2} are depicted with solid lines, whereas dashed lines represent the power series approximation of \eqref{eq:approx omega1} and \eqref{eq:approx omega2}.

Lower branch emerges from the Larmor's frequency $\gamma H$ and grows parabolically with $k$. Effect of inertia reduces the coefficient at the term quadratic in $k$. Upper branch is simply displaced by $+1/\beta$ and has the similar shape.
Imaginary parts of the roots $\omega_{1,2}^{\prime\prime}$ represent attenuation of the corresponding magnetization dynamics in time as $\propto\exp \left (\omega_{1,2}^{\prime\prime}t \right)$, and therefore they must be negative.  In the frequency domain they characterize the width $\Delta f=\left |  {\omega}'' \right |/\pi$ (FWHM) of the Lorentzian spectral line. 

\begin{eqnarray}
\omega _1^{\prime\prime} &\approx& -\alpha\gamma\big[D{k^2} + H - 6\beta {\gamma^2} H D {k^2} + ...\big]\label{eq:approx omega1_imag}\\ 
\omega _2^{\prime\prime} &\approx& -\frac{\alpha}{\beta} -\omega_1^{\prime\prime}\label{eq:approx omega2_imag}  
\end{eqnarray}

Note that in the limiting case of \eqref{eq:approx omega1_imag}, \eqref{eq:approx omega2_imag} field and exchange stiffness have opposite effects on the attenuation of the two magnon branches: they increase the attenuation in the lower branch $\omega _{1}$ and decrease it for the inertial branch  $\omega _{1}$. In the other limiting case of large field and large $\mathit k$ attenuation of both branches tends to $\exp \left ( -\frac{\alpha}{2\beta }t \right )$.
The damping for both branches appears to be naturally proportional to the Gilbert damping parameter $\alpha$. A conspicuous decrease of nutation linewidth $\omega _2^{\prime\prime}(k=0)$ with growing $\alpha$ reported by Cherkasski et al. \cite{cherkasskii2020nutation} roots back to the parametrization of the nutation phenomenon in terms of a product $\alpha\tau$, where $\tau$ denotes the characteristic nutation lifetime. Within this parametrization, a variation of $\alpha$, while keeping $\tau$ constant, leads to the simultaneous decrease of the nutation frequency $1/\beta=1/(\alpha\tau)$ rendering the analysis of damping extremely difficult. An alternative notation in terms of $\alpha$ and $\beta$, introduced in this paper, resolves this problem and allows for an independent investigation of inertial and damping effects.\\
Another parameter, which characterizes the resonant spectral line centered at frequency $f_0$, is its quality factor defined as $Q= f_0 / \Delta f ={\omega}'/\left ( 2{\omega}'' \right )$. As can be seen from equations \eqref{eq:exact omega1} and \eqref{eq:exact omega2}, Q-factors for both branches coincide within the accuracy of $\sim {(\omega \alpha)}^2$. Dependence of the Q-factor on the wavenumber $k$ looks counterintuitive in that it essentially grows with $k$. Assuming for simplicity $H=0$, for small $k$ the Q-factor can be approximated by expansion of ${\omega}'$ and ${\omega}''$ in the power series in $k$, which results in:

\begin{equation}
  Q(k) \sim {\frac{1}{2 \alpha }\frac{1+\gamma \beta D {k^2}}{1-\gamma \beta D {k^2}}+...}
\sim {\frac{1}{2 \alpha}\left ( 1+2 \gamma \beta D {k^2}  \right )}  
\label{eq:Q approx}
\end{equation}

\begin{figure}
	{\footnotesize{} \centering
 \includegraphics[width=1.0\columnwidth]{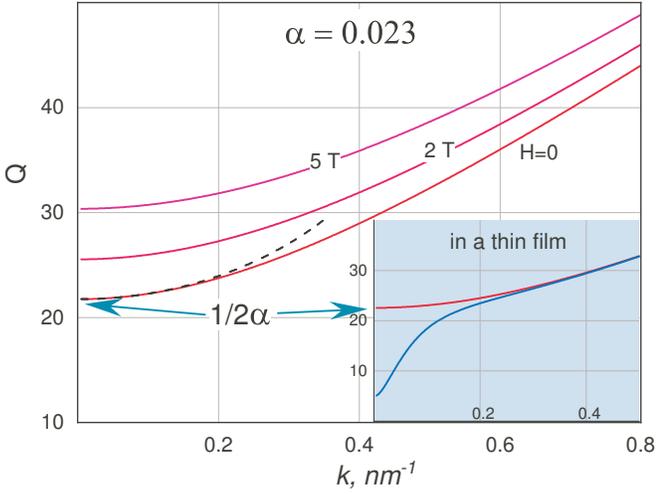}
 \caption{\label{fig:Qfactor}Q-factor dependencies on $k$ for different values of the external field. Dashed line shows the quadratic in $k$ approximation for $H = 0$. Inset shows Q-factors for the ordinary magnon (blue curve) and inertial magnon (red curve) in a thin film, $H=0$}}
\end{figure}

Exact values for the Q-factor in comparison to the estimate of \eqref{eq:Q approx} are shown in Fig. ~\ref{fig:Qfactor} for the external field ranging from 0 to 5 T. \\

Field effect for small $k$ can be approximated as $Q \sim 1/(2\alpha)(1+2\gamma \beta H)$.

\section{Inertial exchange magnons in samples with cylindrical symmetry}
Examples of such bodies are disks, wires, infinite plates and films. Axial symmetry about the $z$-axis retains the diagonal form of $\widehat{N}$ with the diagonal elements satisfying the following conditions: $N_{x}=N_{y}$ and $N_{x}+N_{y}+N_{z}=1$.  As a result, components of the matrix $\widehat{A}$ given by \eqref{eq:A_ij} acquire terms proportional to $\gamma M_S$. Lack of symmetry makes the magnon propagation dependent on the orientation of vectors $\mathbf{m}_0$ and $\mathbf{H}$ with respect to the $\mathit{z}$-axis. We consider two limiting cases: collinear arrangement with $\mathbf{m}_0$ parallel to the axis of symmetry ($\Theta = \Psi = 0$ in Fig.~\ref{fig:Arrangement}); and orthogonal arrangement with $\mathbf{m}_0$ parallel to the $x$-axis and $\Theta = \Psi = \pi /2$.  
In the collinear case the demagnetizing field acts simply against the external field, hence the secular equation remains similar to \eqref{eq:unbounded det}, but with field $H$ substituted with the reduced field ${H}'= H-\xi M_S$:

\begin{eqnarray}
\left(\gamma {H}' + \gamma D k^2 - \beta\omega^2 - i\alpha\omega + \omega  \right)\times\label{eq:film det}\\ \times\left( \gamma {H}' + \gamma D k^2 - \beta\omega^2 - i\alpha\omega - \omega  \right) = 0\nonumber
\end{eqnarray}

where $\xi=N_z-N_x=N_z-N_y$ characterizes the shape effect on demagnetizing, so that in an infinite wire $\xi=-1/2$, in the spherical symmetric (or unbounded) body $\xi=0$  and in the infinite film $\xi = 1$.
Correspondingly the roots to \eqref{eq:film det} are similar to ones given in \eqref{eq:exact omega1} and \eqref{eq:exact omega2} with modified field:

\begin{equation}
{\omega _1} = \frac{1}{{2\beta }}\left( { - 1 - i\alpha  + \sqrt {4\gamma \beta (D{k^2}+{H}') + {{(1 + i\alpha )}^2}} } \right)
\label{eq:film omega1}
\end{equation}

\begin{equation}
{\omega_2} = \frac{1}{{2\beta }}\left( {1-i\alpha  + \sqrt {4\gamma \beta (D{k^2}+{H}') + {{(1-i\alpha )}^2}} } \right)
\label{eq:film omega2}
\end{equation}

At the low-$\mathit k$ limit the lower branch roughly tends to the Larmor's frequency $\gamma (H-\xi M_S)$ and the upper branch limit is $1/\beta +\gamma (H-\xi M_S)$, which is similar to the case of spherical symmetry, but with modified field. 
In the orthogonal configuration with $\mathbf{m}_0$ and $\mathbf H$ perpendicular to the axis of symmetry and to the magnon propagation direction, roots of the determinant \eqref{eq:determ} generally cannot be found in an analytical form. Therefore we first consider an approximate solutions, and then describe briefly the numeric algorithm for obtaining the dispersion curves. 
By neglecting the Gilbert attenuation ($\alpha =0$), the approximate solutions to \eqref{eq:determ} for the in-plane magnetization and field can be found in a concise analytical form: 

\begin{eqnarray}
\omega _{1,2}= \frac{1}{{\beta \sqrt 2 }}\{ 2\gamma \beta (D{k^2} + H) + \gamma \beta \xi {M_s} +1\nonumber \\ \mp\sqrt {4\gamma\beta(D{k^2} + H)+(\gamma \beta \xi {M_s} + 1)^2}\}^{1/2}  \label{eq:Demag omega}
\end{eqnarray}

Here indices  1 and 2 denote the frequencies of the conventional and inertial magnons respectively, sign ‘-‘ prior to the square root in (\ref{eq:Demag omega}) corresponds to the lower branch $\omega _{1}$; sign ‘+’ denotes the inertial branch $\omega _{2}$.
Numerical procedure for building the dispersion relations of the magnonic modes for nonzero $\alpha$ or arbitrary orientation of the external field H starts with calculation of the stationary equilibrium magnetization $\mathbf{m}_0=(m_x,m_y,m_z)$. This can be done by solving \eqref{eq:ILLG} in its stationary form, {\it i.e.} with all time derivatives set zero. In a thin film, for example, quantities $\mathit H_{i}$ and $\mathit m_{j}$ are related by $M_{S}m_{x}m_{z}+H_{x}m_{z}-H_{z}m_{x}=0$. Thus obtained stationary magnetization components are then substituted into \eqref{eq:A_ij} and \eqref{eq:determ}. At some fixed small $k$ the determinant \eqref{eq:determ} as a function of complex-valued $\omega$ possesses two minima, which correspond to the FMR and nutational branches. Their exact locations can be evaluated by a numerical routine which minimizes the absolute value of the determinant \eqref{eq:determ} in the vicinity of the guess values for those branches, for example given by equations \eqref{eq:approx omega1 Demag} and \eqref{eq:approx omega2 Demag}. Then we give $k$ a small increment and repeat the extremum search using the $\omega$s obtained at the previous step as guess values, and so on. As a result, calculated values for $\omega_1$ and $\omega_2$ follow the dispersion curves of both branches. Note that for nonzero $\alpha$ roots $\omega_{1,2}$ possess imaginary parts, which determine the line width and Q-factor for each mode.   
Let us consider the magnetization behavior in a thin film in more detail. For this geometry $\widehat{N}$ possesses the only nonzero component $N_z=1$, and correspondingly $\xi =1$. In the small-$k$ limit, the lower branch approaches the Kittel's frequency $\omega_{FMR}=\gamma \sqrt{H\left (H+M_{S}  \right )}$ from below as $\beta ,k\rightarrow 0$:

\begin{equation}
\omega_{1}\approx \omega_{FMR}-\frac{1}{2}\gamma\omega_{FMR}\left ( 2H+M_{S} \right ) \beta+... 
\label{eq:approx omega1 Demag}
\end{equation}

Effect of the demagnetizing field on the inertial branch is exhibited by an upward shift by $\frac{1}{2}\gamma M_{S}$; whereas effect of inertia is opposite:

\begin{equation}
\omega _{2}\approx \frac{1}{\beta }+\gamma H+\frac{1}{2}\gamma M_{S}-\left ( \omega _{FMR}^{2}+\frac{1}{8}\gamma ^{2}M_{S}^{2} \right )\beta +...
\label{eq:approx omega2 Demag}
\end{equation}

\begin{figure}
	{\footnotesize{} \centering
 \includegraphics[width=1.0\columnwidth]{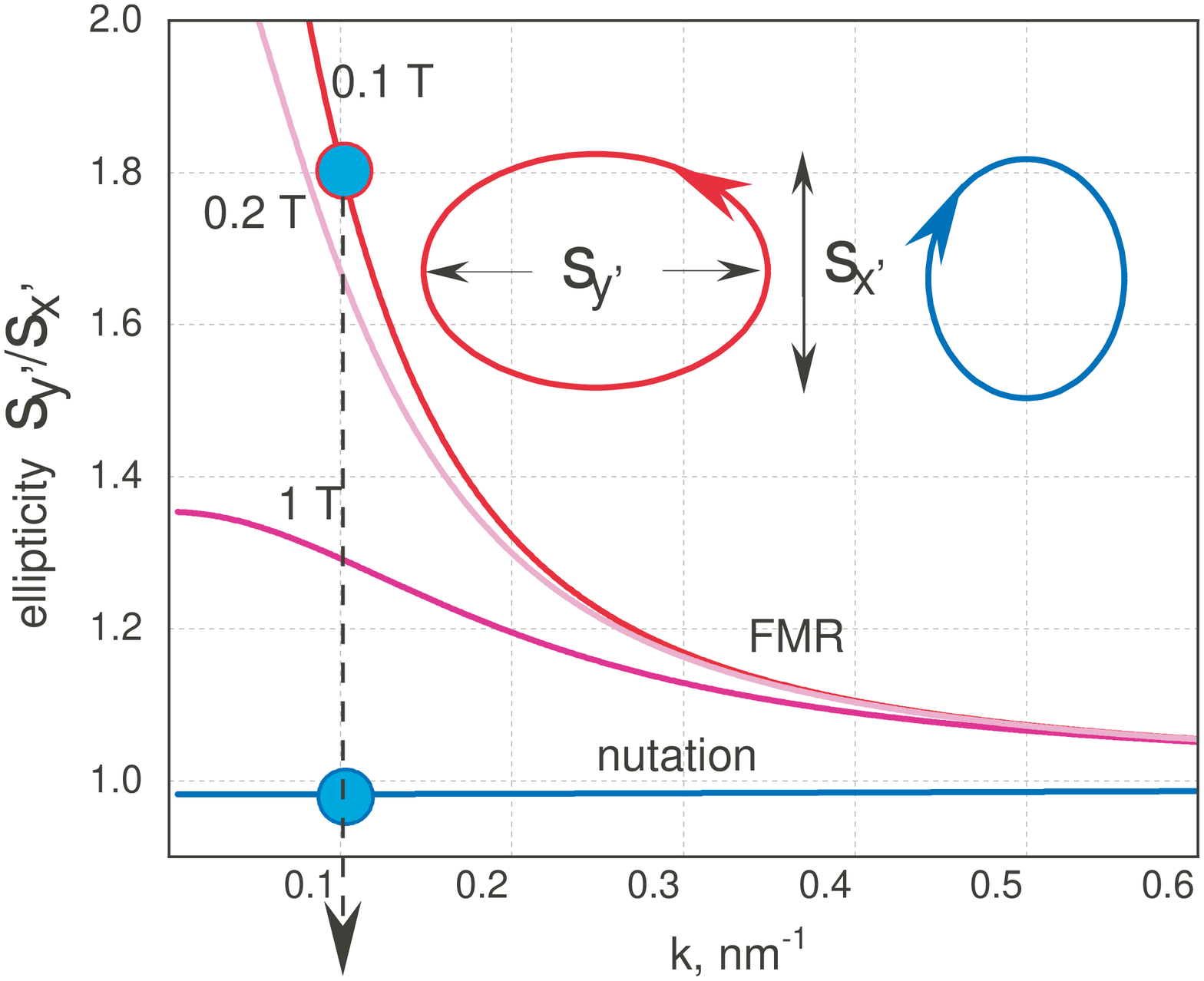}
  \includegraphics[width=0.85\columnwidth]{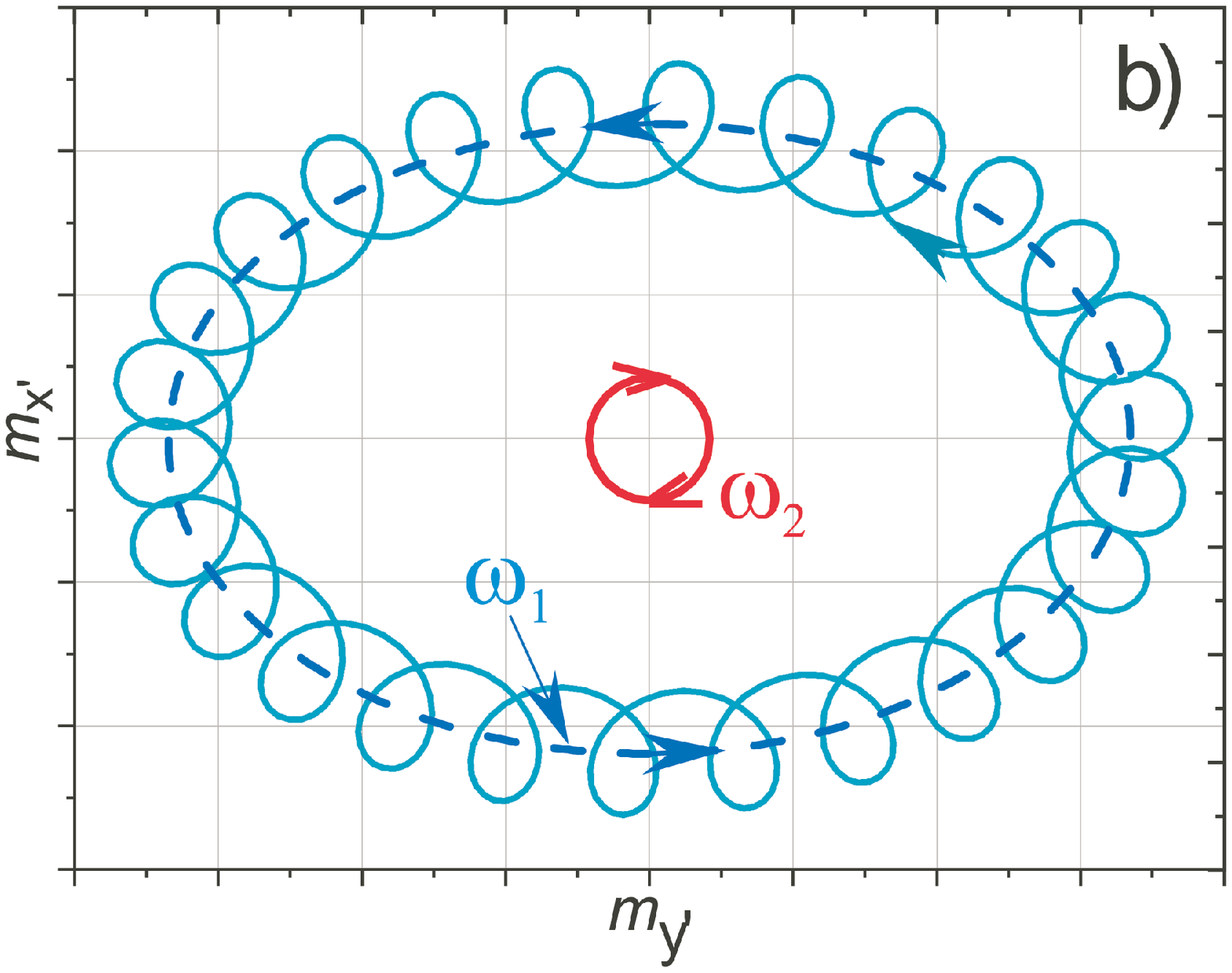}
 \caption{\label{fig:ellipticity} Ratio of the polarization axes for the external field of 0.1T, 0.2T, and 1T. $\mathbf H$ and $\mathbf{m}_0$ are parallel to the inward normal to the figure plane. }}
\end{figure}

For the orthogonal configuration, when both $\mathbf{m}_0$ and $\mathbf{H}$ lie in the film plane, we can estimate the trajectories of the magnetization dynamics of both modes for small $k$ and small $\alpha$. For each root given by \eqref{eq:Demag omega} we solve the homogeneous equation for perturbations $s$ \eqref{eq:system}. Normalization of the solutions can be chosen in an arbitrary way,here for simplicity we define $s_{z} =1$. In orthogonal geometry $s_x$ component is obviously negligible or equals zero, so the system reduces to two equations in $s_y$ and $s_z$. Results shown as a power series expansion for small inertia $\beta \omega \ll 1$ for the precession:

\begin{equation}
s_p=\begin{pmatrix}
0\\ 
-i\sqrt{1+\frac{\xi M_S}{D{k}^2+H}}\left (1+\frac{\beta \gamma \xi M_S}{2}\right )\\ 
1\end{pmatrix}\exp(-i\omega_1 t)
\label{eq:traj precession}
\end{equation}

and nutation:

\begin{equation}
s_n=\begin{pmatrix}
0\\ 
i\left (1-\beta \gamma \xi M_S /2 \right )\\ 
1\end{pmatrix} \exp(-i\omega_2 t)
\label{eq:traj nutation}
\end{equation}

with the parameter of anisotropy $\xi = 1$ for a thin film normal to the $z-$ axis and $\xi = 1/2$ for a thin wire spread along the $z-$ axis.
Both perturbations exhibit elliptical polarization within the $(x^\prime,y^\prime)$ plane. Precession trajectory is deformed by the demagnetizing effect so that the $y$-axis of the ellipse is stretched with the $\sqrt{1+M_S/H}$ factor due to demagnetizing effect, and in addition on account of inertial effect. On the contrary, the nutation ellipse is squeezed along the $y$-axis proportionally to the inertial parameter $\beta$. Ellipticity of the lower branch depends on the external field \eqref{eq:traj precession}, whereas that of the upper branch in this approximation shows no dependence on the field. Signs of $s_y$ components are opposite for nutation and precession, this indicates that they are rotating in the opposite directions. Exact polarizations can be found numerically for a reasonable set of material parameters and fields, as is shown in Fig.~\ref{fig:ellipticity}. 

\section{Excitation mechanisms of inertial exchange magnons}
The only experimental evidence of inertial effects in ferromagnets has been reported for $k=0$ nutation mangons in Py-thin films resonantly excited with a  magnetic field of an intense quasi-monochromatic THz pulse \cite{neeraj2021inertial}. In order to excite $k\neq 0$ exchange magnon modes one would need to have either spatially localized and instantaneous stimuli \cite{razdolski2017nanoscale} or any other source of effective magnetic field characterized by spectral and spatial overlap with investigated magnon modes. The letter can be provided through ultrashort large-amplitude acoustic pulses \cite{Temnov2013NatureComm,Temnov2016} producing effective magneto-elastic fields rapidly varying in time and space \cite{BesseJMMM}. Acoustic pulses propagating through a thin ferromagnetic sample at an acoustic velocity $v$ are quantified by a linearized dispersion relation $\omega_{ac}=vk$.  Crossing between acoustic and magnon brunches, i.e. satisfying the phonon-magnon phase-matching condition, usually facilitates the acoustic excitation of magnetization dynamics \cite{janusonis2016ultrafast,chang2017parametric}. A question arises under which conditions the crossing between dispersion curves for longitudinal phonons and inertial magnons can occur. Whereas for realistic magnetic fields the acoustic dispersion always intersects the lower FMR-branch at a frequency close to FMR frequency \cite{BesseJMMM}, the crossing of the upper nutation brunch is less obvious.  

\begin{figure}[htb!]
	{\footnotesize{} \centering
 \includegraphics[width=1.0\columnwidth]{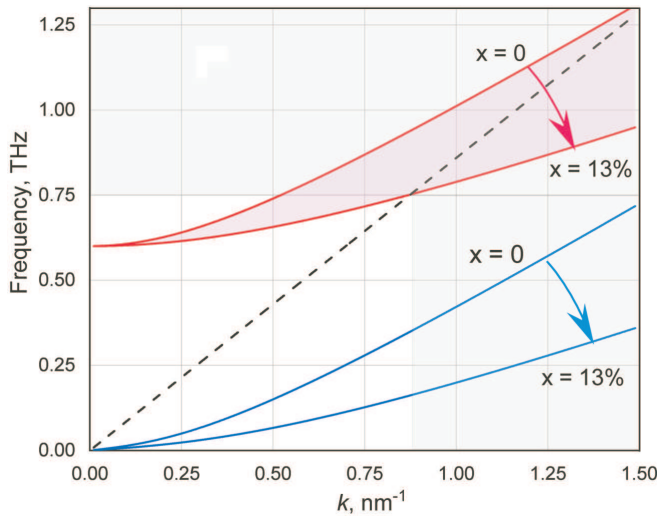} 
 \caption{\label{fig:Demag effect} 
 The magneto-acoustic phase matching condition for nutation magnons can be tuned vie the reduction of exchange stiffness in Gd-doped Py samples. Gd concentration x varies from 0 to 13\%. The dashed line displays the acoustic dispersion relation $\omega _{ac}/(2\pi)$.  Magneto-elastic coupling with inertial mangon is efficient when the dashed line lies within the pink tinted area. Material parameters are taken from Ref.~\cite{salikhov2019gilbert} and $\beta$=0.276~ps.}}
\end{figure}

It is possible to quantify the criterion for magneto-elastic crossing with nutation magnons analytically. To do that we note that for larger wavenumbers $\mathit k$ satisfying $Dk^{2}\gg H, M_{S}$ the exchange term plays the dominant role and the asymptotic behaviour for both branches becomes linear in $\mathit k$:

\begin{equation}
\omega _{1,2}\approx \mp \frac{1}{2\beta }+k\sqrt{\frac{\gamma D}{\beta }}\,.
\label{eq:approx great k}
\end{equation}

It follows from (\ref{eq:approx great k}) that the condition for the nutation magnon branch to intersect the acoustical dispersion relation  $\omega _{ac}(k)$, requires the asymptotic slope of $\omega _{2}(k)$ to be smaller than the acoustic velocity $v$:

\begin{equation}
\sqrt{\frac{\gamma D}{\beta }}<v\,.
\label{eq:acoust couppl}
\end{equation}

This expression shows that for a given $\beta$ the magneto-elastic crossing is facilitated by small exchange stiffness $D$ and small acoustic velocity. This approximate analysis breaks down for acoustic frequencies in above-THz spectral range, where the acoustic dispersion starts deviating from its linear approximation.  

Figure 5 highlights the remarkable role of exchange stiffness to achieve the dispersion crossing between nutation magnons and longitudinal acoustic phonons. Doping Py thin films with Gadolinium has been shown to gradually reduce the exchange stiffness upon Gd-concentration from 300 to 100~[meV$\cdot$\text{\normalfont\AA}$^2$] \cite{salikhov2019gilbert}. For a fixed value of inertial parameter $\beta$=0.276~ps, nutation magnons for pure Py samples do not display any crossing with acoustic phonons within the displayed range of k-vectors but the Gd-doped Py with 13\% Gd concentration does. The nutation magnon-phonon crossing point occurs at 0.75~THz frequency and $k=0.85$~nm$^{-1}$ (magnon wavelength of approximately 5~nm), i.e. magnon parameters readily accessible in ultrafast magneto-optical experiments \cite{razdolski2017nanoscale}.      

\section{Conclusions}
 In this paper we have theoretically studied exchange inertial magnons in ferromagnetic samples of different shapes under the action of an external magnetic field. The parametrization of magnetization dynamics in terms of two independent parameters, the  Gilbert damping $\alpha$ and the inertial time $\beta$, allows for unambiguous discrimination between the inertial and damping effects as well as their impact on both branches of magnon dispersion. Inertial effects are found to strongly effect not only the frequencies (magnon eigenvalues) of both branches but also result in a monotonous increase of the Q-factor as a function of the external magnetic field and magnon $k$-vector. The two magnon branches are found to precess in opposite directions along the elliptical trajectories with perpendicularly oriented long axis of the ellipses (magnon eigenvectors). Their ellipticity is found to depend on the components of the demagnetizing tensor. An analytical criterion for the existence of phase-matched magneto-elastic excitation of nutation magnons has been derived and illustrated for Gd-doped permalloy samples with tunable exchange stiffness.      
 
\begin{acknowledgments}
Financial support by Russian Basic Research Foundation
(Grant No. 19-02-00682) is gratefully acknowledged.
\end{acknowledgments}

%\bibliographystyle{unsrt}
%\bibliography{Biblio}
% 

\end{document}